\documentclass[
    final          
  ]{aipproc}

\layoutstyle{6x9}

\begin{document}

\title{Dark Energy from Strings}

\author{Paul H. Frampton}
{address={Department of Physics and Astronomy, University of North Carolina, Chapel Hill, NC 27599-3255}}

\begin{abstract}
A long-standing problem of theoretical physics is the exceptionally small
value of the cosmological constant $\Lambda \sim 10^{-120}$ measured
in natural Planckian units.
Here we derive this tiny number from a toroidal string cosmology
based on closed strings. In this picture the dark energy
arises from the correlation between momentum and winding modes that
for short distances has an exponential fall-off with increasing values of the
momenta.The freeze-out by the expansion of the background universe for
these transplanckian modes may be interpreted as a frozen condensate
of the closed-string modes
in the three non-compactified spatial dimensions.
\end{abstract}

\maketitle


\section{Introduction}

This talk is based on a paper with Bastero-Gil and Mersini, necessarily
shortened to fit the space available, so I refer to it\cite{BFM}
for more detail.

In this work we will attempt to make a quantitative argument
about the origin of dark energy from string theory.
The transition from string theory to conventional cosmology
is of importance not only to theoretical physics in general but
to inflationary cosmology in particular. Corrections to
short distance physics due to the nonlocal nature of strings
contribute to dark energy. The possibility to detect their
signature observationally  is very intriguing.
In Ref. \cite{paper1} it was shown that a nonlinear dispersion function
modifying the frequency of the transplanckian perturbation modes \cite{tp}
can produce  the right contribution to the dark energy of the universe
\cite{dark}. The  physics mechanism that gave rise to dark energy was
the freeze-out of these ultralow frequency modes by the expansion of
the background universe.
Superstring duality \cite{tduality} was invoked to justify the dispersion
function. This work attempts to carry out this derivation.

In Section 2 we review some preliminaries of the
Friedman-Robertson-Walker (FRW) cosmological solutions
found for string theory in a D-dimensional torus
\cite{vafa,mueller,tvafa,tseytlin}.
The quantum hamiltonian from closed string theory obtained in
\cite{tseytlin2} by using the correspondence principle between string
and quantum operators, is reviewed in Section 3. Although the
background is an FRW universe, it is globally nontrivial 
and thus it allows two types of quantum string field
configurations, twisted and untwisted fields.

Based on the equivalence between Euclidean path integral
and statistical partition functions, we perform in Section 4 the
calculation of a  coarse-grained effective action \cite{blhu,huang}
for the momentum and
winding modes of  the system described in Section 2 for the case of 3
expanding spatial dimensions $R$ in the $T^D$ toroidal
topology.
The string scale is taken as the natural UV lattice cutoff scale of the
theory. The renormalization
group equations (RGE) of the coupling constants for the winding and
momentum modes describe the evolution from early to late times of
their entanglement.
Based on T-duality the whole spectrum is
obtained by exchanging momentum to winding modes
and $R \to R^{-1}$. Their coupling is strong when the radius of the
torus $R$ is of the same order as the string scale $\sqrt{\alpha^\prime}$,
i.e. during
the phase transition from a winding dominated universe to a
momentum mode dominated universe. Due
to the expanding background, we have a non-equilibrium
dynamics
and calculate the effective action by splitting our modes into the open
system degrees of freedom (low energy modes, mainly momentum modes) and the
environment degrees of freedom (high energy modes, mainly winding modes).
The coarse-graining is performed by integrating out the environmental
degrees of freedom.
The scale factor $a(t)=R(t)$ serves as the collective coordinate that
describes the order parameter for the environment degrees of freedom.
 The effective action calculated
in this way contains the influence of the environment at all times in a
systematic way and the coarse graining process encodes the dispersion
function and corrections to short distance physics due to the
correlation between the two types of modes in the system and
environment. This procedure results in the
RGEs for the coupling constants that offers information about their
running to trivial and nontrivial fixed points at early and late times,
therefore the flow of one family of lagrangians (string theory phase)
to another family of lagrangians at late times (conventional 3+1 quantum
theory). Results of this non-equilibrium phase transition are summarised in
Section 5 with a discussion about the possibility of their observational
signatures through the equation of state of the frozen short distance modes.
In this section we also briefly touch upon the issue of the two field
configurations in a globally nontrivial topology and the instabilities
in the theory arising from their interaction. A detailed summary of
the main coarse-graining formulas and procedure \cite{blhu} needed in
Section 4,
are attached in the Appendix. In essence, the dark energy arises from the
study of the UV behavior of the correlations with environmental modes.

\section{Toroidal String Cosmology.}

We consider the string cosmological scenario proposed by
Brandenberger and Vafa\footnote{Herein referred to as the BV model.}
\cite{vafa,tvafa,others}.
Strings propagate in compact space, a box with
D spatial dimensions and periodic boundary conditions, the $T^D$ torus.
It was argued that \cite{vafa} a thermodynamic description  of the
strings with positive specific heat, is well defined only when all the
spatial dimensions
are compact.

Let us begin with the Universe placed in a $T^D$ box
with a size of the order of the string scale, that we are taking to
be the Planck scale. In such a
space, string states also contain winding modes, which are
characteristic of having an extended object like a string,
``winding'' around the compact spatial dimension, besides the usual
momentum modes, and oscillator modes with energy independent of the
size of the box. The
energy of the winding modes
increases with the size of the box as $w R$, while the energy of the
momentum modes decreases as $m/R$. The spectrum is symmetric under the
exchange $R \leftrightarrow 1/R$ and $m
\leftrightarrow w$. This symmetry known as T-duality \cite{tduality} is not
only a symmetry of the spectrum but of the theory.

The BV model \cite{vafa} argues that if the Universe expands
adiabatically in more than 3 spatial dimensions, it would
not be possible to maintain the winding modes in thermal
equilibrium. As their energy density grows with the radius,
their number would have to decrease, for example through annihilation
processes. But typically strings do not meet in more than 3 spatial
dimensions  and do not interact with each other; therefore the
winding modes fall out of equilibrium \cite{numerical}. In
summary, their growing energy density will tend to slow down the expansion
of the universe and eventually stop it.
But if the Universe starts to contract, the dual
scenario of the momentum modes opposing contraction would take place
and the Universe
may oscillate between expanding/contracting eras. In what follows we
use this argument of \cite{vafa} to justify the assumption that only
$D=3$ dimensions of the $T^D$ torus will expand to create an FRW
universe.

Cosmological solutions for an arbitrary number of anisotropic
toroidal spatial dimensions $T^D$ were found by Mueller in
\cite{mueller}. He studied the cosmology of bosonic strings
propagating in the background defined by a time-dependent dilaton field,
$\Phi(t)$, and space-time metric
\begin{equation}
ds^2_d=G_{\mu\nu}(X) dX^\mu dX^\nu=-dt^2+ \sum_{i=1}^D 4 \pi R_i^2(t)dX_i^2\,,
\label{metric}
\end{equation}
The radii of the torus, $R_i(t)$, become the time-dependent
scale-factors, and the spacetime dimensions is $d=1+D$. The equations
of motion of the bosonic string in
background fields are obtained from the following action\footnote{The
antisymmetric tensor field is taken to be zero.} \cite{action}
\begin{equation}
I= \frac{1}{4 \pi \alpha^\prime} \int d^2 \sigma \sqrt{g} \left[
g^{mn} G_{\mu\nu} (X) \partial_m X^\mu \partial_n X^{\nu} + \frac{1}{2}
\alpha^\prime \Phi R^{(2)} \right] \,,
\end{equation}
where $g_{mn}$ is the two-dimensional world-sheet metric, and
$R^{(2)}$ the world-sheet scalar curvature. The background field
equations are obtained by imposing the condition that the theory be free from
Weyl anomalies. To lowest order in perturbation theory this leads
to the equations:
\begin{eqnarray}
\beta_{\mu\nu}^G &=& R_{\mu\nu} + \nabla_\mu \nabla_\nu \Phi =0 \,,\\
\beta^\Phi &=& \frac{d-26}{3 \alpha^\prime} - R + (\nabla \Phi)^2 -2
\nabla^2 \Phi =0 \,,
\end{eqnarray}
Using the metric given in Eq. (\ref{metric}), they reduce to:
\begin{eqnarray}
\ddot{\Phi} - \sum{i} \frac{\ddot{R}_i}{R_i} &=&0 \,, \label{bg1}\\
 \frac{\ddot{R}_i}{R_i} + \sum_{j\neq i} \frac{\dot{R}_i
 \dot{R}_j}{R_i R_j} - \dot{\Phi} \frac{\dot{R}_i}{R_i} &=&0 \,,\\
\ddot{\Phi} - \frac{1}{2} \dot{\Phi}^2 + \sum_{i<j} \frac{\dot{R}_i
 \dot{R}_j}{R_i R_j} &=& \frac{ d-26}{3 \alpha^\prime} \,. \label{bg3}
\end{eqnarray}
When $D=25$, the solutions obtained in \cite{mueller} are:
\begin{eqnarray}
e^{-\Phi(t)} &\propto& t^p \,, \\
R_i(t) &\propto& t^{p_i} \,,
\end{eqnarray}
with the constraints,
\begin{equation}
\sum_{i=1}^D p_i^2 = 1 \,,\;\;\;\;\;
\sum_{i=1}^D p_i = 1 - p \,.
\label{muellers}
\end{equation}
Note that these solutions are found in the absence of matter sources.
In general the backreaction of the matter action of the strings
in $T^D$ alters the solutions for the background geometry\footnote{See
\cite{tvafa,tseytlin}
and references therein for the geometry solutions in the presence of a
matter action. Inclusion of matter
sources alters the solutions of \cite{mueller}
due to the backreaction of the winding modes, such that
the scale factor approaches asymptotically a constant
value at late times
.}. It is clear that we can have an arbitrary number of compact
spatial dimensions
$D_c$ with $p_i < 0$, that are decreasing with time
\footnote{We do not address the concern that the time dependence of
the compactified $R_i$
endangers the constancy of the dimensionless parameters in the $D=3$
theory.}, and $D-D_c$ expanding spatial
dimensions with $p_i >0$. Among the many solutions found in
\cite{mueller} we select the solution $D-D_C=3$  that although it is
not unique it is justified by the BV argument.
The assumption that our Universe is expanding in only 3 spatial
dimensions, with the remaining $D-3$ being small and compact, as well
as considering a constant dilaton field\footnote{The authors of \cite{tvafa}
argued that a constant dilaton background may not be consistent with
a $high$ temperature phase of strings thermodynamics.}
($p=0$), are consistent with Mueller's solutions
Eqs. (\ref{muellers}). The issue of stabilising the dilaton is beyond
the scope of this
paper, and we assume that the dilaton has acquired a mass
and become stable at some fixed value. It is also assumed
that the backreaction of the matter string sources
on the backround geometry is small enough such that the deviations from the
FRW metric, Eq. (\ref{muellers}), can be neglected.

Due to the toroidal string cosmology, the three expanding dimensions
contain both types of modes: $momentum$ and $winding$, propagating
in the 3+1 FRW space-time. The
number of winding modes at each
stage of the evolution of the Universe is determined by the
dynamics of the background. In the next section, we touch base with
quantum field theory through correspondence principle between string
and quantum operators, in order to use coarse
graining techniques for studying the influence of the winding modes
on the momentum modes as the Universe expands.

\section{Quantum Hamiltonian from Closed String Theory. }

Let us consider BV model \cite{vafa} of a
D-dimensional anisotropic torus with radius $\bar{R_i}$, by including
the dynamics of both modes: momentum modes, $p_{1,i}=m/\bar R_i$
(where $m$ is the
wavenumber), and winding modes with momenta $p_{2,i}=w\bar
R_i/\alpha^\prime$. The
dimensionless quantity for the radius is
$R_i=\bar{R_i}/\sqrt{\alpha^\prime}$, where $\alpha^\prime$ is
the string scale. Based on the arguments reviewed in Section 2, we
{\it choose} a
cosmology with three toroidal radii equal and large $R \gg 1$ in
units of the string or Planckian scale, with
the other $(D - 3)$
toroidal radii equal and small $R_C \ll 1$. Here the subscript $C$
refers to compactified dimensions.
Then,  $R(t)$ becomes the scale factor
for the 3+1 metric in conventional FRW (Friedman Robertson Walker)
cosmology $R(t)=a(t)$, while $R_C$ corresponds to
the radius, in this factorizable metric, of the $D-3$ compact
dimensions $z_j$ that decrease with time,
\begin{equation}
d s_D^2 = -dt^2 + 4 \pi R^2(t) dx_i^2 + 4 \pi R^2_{C}(t) dz_j^2 =
a(\eta)^2 [- d\eta^2 + dy^2] + d s_{D-3}^2\,.
\label{metric2}
\end{equation}
Using the string
toroidal solution of \cite{mueller} the time-dependence of these radii is:
\begin{equation}
R(t) = \alpha_U t^{p_U}
\end{equation}
\begin{equation}
R_C(t) = \alpha_C t^{p_C}
\end{equation}
The solutions in Ref. \cite{mueller} show that $p_U$ and $p_C$ depend
on the dimensionality $D$ in an interesting way. There is a plethora
of possible solutions but if we assume, for example, that the dilaton
is time-independent and the compactification is isotropic we find that
for $4 \leq D < \infty$, then $0.5 \leq p_U < 1/\sqrt{3} \simeq
0.577$. Let us
take $D=4$ where the scale factor behaves as a radiation-dominated
universe; if, in fact, $D \geq 5$ we can assume that the $D-4$
additional dimensions have $p_C^{'} \ll p_C$ to achieve the same
result. In this case, $p_C = -0.5$. Here we do not, however, need to
specialise to a particular solution.

What we have in mind for the dark energy is the correlation of momentum to
{\it winding} string modes.
The question is, given the well-known form for the kinetic energy of these
strings, {\it e.g.} \cite{kikkawa}, how to describe best the interaction
between the winding and momentum modes. Some aspects are addressed in
\cite{kikkawa} which focuses on the smallness of temperature $(T/T_H)$.
For temperature $T$ very much below the
Hagedorn or string temperature $T_H$ we expect  that
only very small winding numbers $w_i = 0$ or $1$ in the compact
dimensions are of any significance \cite{kikkawa}. Similar arguments
apply to the momentum modes $m_i$ for the time-reversed case.

Let us consider the small parameter $\delta(t)$, taken to be:
\begin{equation}
\delta = \frac{R_C}{R} \sim t^{p_C - p_U}
\end{equation}
For the case $D=3$ (d=4), for example $\delta \sim t^{-1} \sim (T/T_H)^2$
and is an extremely small number ($\sim 10^{-60}$) at present. The
point is that in the $\delta \rightarrow 0$ limit these modes are
in separate spaces and for very small $\delta$ are therefore expected
to be highly restricted. The compactified dimensions can be integrated
out, and we are left with the momentum and winding modes in the
remaining $D=3$ spatial dimensions.

The partition function for this system was  calculated, from first
principles, by summing up over their momenta in \cite{kikkawa}:
\begin{equation}
Z= \sum_\sigma e^{-n_\sigma \epsilon_\sigma}\,,
\end{equation}
where $n_{\sigma}$ is the number of strings in
state $\sigma$ with energy $\epsilon_{\sigma}$
\begin{equation}
\epsilon_\sigma=p_0=\sqrt{\left(\frac{m}{R}\right)^2 + (w R)^2 + N +
\tilde{N} -2 } \,,
\label{epsilon}
\end{equation}
and $\sigma$ counts over $(m,\,w)$, with the constraint $N-\tilde{N}=m
w$ for closed strings where $N$ and $\tilde{N}$ are the
sums over the left- and right- mover string excitations, respectively.
By now, in Eq.(\ref{epsilon}),
we are considering only the large 3 spatial dimensions.
The string state can also be described by
its left and right momenta, $k_L = p_1+p_2$, $k_R=p_1 - p_2$. The string
state for left and right modes can be
expanded in terms of the creation and annihilation operators
$\alpha_m$, $\tilde\alpha_n$, with higher excitation string states
given by $N=\sum_{n=1}^{\infty} \alpha_{-n}\alpha_n$ (similarly
for $\tilde{N}$),  and string energy
$L_0 + \tilde{L}_0 = p_1^2 +p_2^2 + (N +\tilde{N} -2)/\alpha^\prime$.

We would like to write the
path integral for this configuration in terms of quantum
fields\footnote{Below we use quantum string equations under the assumption
that the dilaton is massive and stable.}. The path integral is
calculated from the hamiltonian density.
In order to use the correspondence between the Euclidean path integral
of the persistence vacuum amplitude $|\langle in | out \rangle|^2$ and
the partition function $Z$, we need to write a hamiltonian density over the
fields in configuration space  in such a way that its Fourier transform
in $k$-space corresponds to the string energy expression Eq. (\ref{epsilon}).

Thus in writing a Coarse-Grained Effective Action (CGEA), the kinetic terms
are unambiguous while for the interaction terms we must appeal to simplicity
and the requirement of T-duality. Closed-string field theory
 provides guidance, since in {\it e.g.} \cite{belopolsky} truncation
at a quartic coupling can be sensible, and this will lead to a CGEA
which is renormalisable and satisfies useful RG equations.

Generally, closed string field theory contains couplings of all non-polynomial
orders. In a semi-classical approximation
we may restrict to genus $g=0$ since the genus $g$ contribution
is proportional to $\hbar^g$ \cite{SB}.

The quantum hamiltonian is in any case known for the
classical string in axi-symmetric or toroidal
backgrounds \cite{tseytlin2}. They explicitly
calculated the quantum hamiltonian and demonstrated the correspondence
principle between the string operators $L_0$, $\widetilde L_0$ and
quantum field operators in the form (in the notation of \cite{tseytlin2})
\begin{eqnarray}
\hat{H}&=&\hat{L}_0+ \hat{\widetilde{L}}_0 = \frac{1}{2}\alpha^\prime
\left( -E^2 + p_a^2 + \frac{1}{2} (Q_+^2 +Q_-^2)\right) + N + \widetilde{N} -2
c_0  \nonumber \\
& & -\alpha^\prime \left[ (q+\beta) Q_+ + \beta E \right]
-\alpha^\prime \left[ (q-\alpha) Q_- + \alpha E \right] J_L \nonumber
\\
& & \frac{1}{2} \alpha^\prime q \left[(q+2 \beta) J_R^2 + (q-2 \alpha)
J_L^2 + 2 (q+\beta -\alpha) J_R J_L \right]
\label{tseytlin}
\end{eqnarray}
\begin{equation}
\hat{L}_0- \hat{\widetilde{L}}_0 = N- \widetilde{N} -mw
\end{equation}
where $J_{R,L}$ are bilinear quadratic operators in terms of creation
and annihilation operators and the higher string
oscillators $N, \tilde N$ contribute the string mass. Therefore the
$J_R^2$ term would be a quartic interaction in terms of creation and
annihilation operators.

This particular solution is for a cylindrical
topology where the uncompactified
$x_1$ and $x_2$ are written in polar coordinates
$x_1 + ix_2 = \rho e^{i \phi}$ and
$x_3$ is also uncompactified (but could be compactified along
with additional similar coordinates), together with time and one
additional compactified dimension $y \subset (0, 2 \pi R)$.
Although an exact solution for the hamiltonian of the string
matter in a toroidal background is not yet known, a
quartic potential energy was advocated and found in \cite{tseytlin} by
arguments similar to those of Eq. (\ref{tseytlin}), for the classical
string and the three string coupling level.
We take this as an indication, in the subsequent
section (if the exact solution were known to all orders),
that an quantum hamiltonian analogous to Eq.(\ref{tseytlin})
for closed strings
on a torus, similarly containing only
quartic terms as suggested by \cite{tseytlin2}, exists
for our present case
of $(T_3) \times (T_{D-3}) \times (time)$ and focus on the
uncompactified 3 spatial dimensions.

The hamiltonian depicted in Eq.(\ref{tseytlin}) is for a static
background,i.e a constant scale factor
 $R(t)$. In the next section, we base our calculation in the
coarse-grained effective action (CGEA) formalism where the
dynamics  of an expanding background is replaced by scaling on a
static background.

Thus Eq.(\ref{tseytlin}) which applies to a static background (as in Eq.(\ref{epsilon}))
can be generalized to a cosmologically-expanding background as in
Eq.(\ref{metric2}) by using this technique of re-scaling, for details
of which see \cite{BFM}.
The result is a dispersion formula characterized by
a dispersed frequency with short distance
modifications contained in a $\tilde{m}_0^2$ term:
\begin{equation}
\tilde{m_0}^2 \stackrel{p \to \infty}{\rightarrow}
\frac{m_0^2}{2 \cosh^2 p \sqrt{\alpha^{\prime}}} \simeq \frac{1}{2}
m_0^2 e^{-2 \sqrt{\alpha^{\prime}} p} \,.
\label{result}
\end{equation}

\section{Dark Energy from Closed String Theory. Discussion}

We argue that closed strings on a toroidal cosmology
lead to a plausible explanation of the dark
energy phenomenon. Although bosonic strings have been used,
it is expected that superstrings will lead to a similar conclusion.
Certainly it is crucial that closed strings are involved because
open strings do not have the same aspect of winding around the torus.

The scale factor of the universe $a(\eta)$ has been used
as a collective coordinate for the environment
degrees of freedom, and as the fundamental scaling parameter in the
coarse-graining.
The choice of a $D=3$ expanding cosmology was chosen phenomenologically. An
argument for this choice in the BV model was presented in
\cite{vafa}, and we believe this argument does provide a possible
justification. It is encouraging that
inclusion of branes gives a similar result \cite{others}.
It has further been assumed that the mass gap $\Delta_p$ can be
safely assumed to be slowly-varying during
our coarse-graining procedure.

We would like to make the reader aware of another subtlety related to the
torus topology of our background. A globally nontrivial topology like
$T^3 \times R^1$ admits two types of quantum field configurations,
twisted and untwisted fields, due
to the periodic and anti-periodic boundary conditions imposed
on the fibre bundle of the manifold. This is a long-standing
problem \cite{ford} that does not have a definite remedy. The problem is
the following: twisted fields can have a negative
two-point function. These fields
interact with each other while preserving the symmetries of the hamiltonian.
Their interaction thus contributes a negative mass squared term
to the effective mass of the untwisted field due to
the negative two-point function of the twisted field and render the
untwisted field unstable. It is often assumed that Nature simple chose
to preserve the untwisted configuration only or forbids their interaction
due to some, as yet unknown, symmetry \cite{laura}.

String theory preserves
Lorentz invariance. This symmetry has been broken for the
open system of our low energy string modes due to the
backreaction from the coarse grained environment.
Their correlation results in our dispersion relation.
If a specific frame must be chosen, it
could be {\it e.g.} the rest frame of the CMB.
The initial condition  is a vacuum state conformally equivalent
to the Minkowki spacetime - the so-called Bunch-Davis vacuum\cite{BD}.
Finally, before summarising we should note that if there are other
modes without the exponential suppression at high $k$, all that we
need is one such mode to lead to the frozen tail comprising the dark energy.

The high wave number behaviour $e^{-ak}$ of
the dispersion relation $\omega(k)$ leads again to the
correct estimate for the dark energy as a fraction $\sim 10^{-120}$ of
the total energy during inflation.
This dark energy is certainly completely stringy because
our derivation depends
on the existence of winding modes, as seen by the role of
the generalised level-matching condition
\[
N - \widetilde{N} = \Sigma_{i}m_iw_i
\]
This correlation between momentum and winding modes leads to the
quantum hamiltonian and
hence to the interpretation of the dark energy
as the weak correlation with the winding mode energy at short distances.
The excitation modes of these correlations with energy less
than the current expansion rate are currently frozen by the
expanding background.

Within string cosmology there has always been the question of the
fate of the winding modes in the uncompactified three spatial dimensions,
whether they combine to a single string per horizon which
wraps around the universe. Our
remedy is intuitively appealing that while the momentum
modes are in evidence as quarks, leptons, gauge bosons, etc.
the winding modes are now
condensed uniformly in the environmental background,
hence with a weak correlation at short distances to the momentum modes,
frozen by the expansion of the FRW universe in the form of the dark energy.

The observed small value $\Lambda \sim 10^{-120}$ in natural units
bserved small value $\Lambda \sim 10^{-120}$ in natural units
has an explanation in the toroidal cosmology of closed strings and thus the
dark energy provides an exciting opportunity to connect string theory to
precision cosmology.
We may argue that numerically the
size of the cosmological constant in the present approach
is a combination of the string scale
and the Hubble expansion rate in the sense that
$\Lambda/M_{Planck}^4 \simeq 10^{-120} \simeq (H_0/M_{Planck})^2$.
Therefore the correct amount
of dark energy obtained by this frequency dispersion
function does not require any fine tuning and relies,
besides a physical mechanism (such as freeze-out),
only on the string scale as the parameter of the theory.
However, our approach does not solve the second puzzle about the dark
energy namely, the coincidence problem for the following reason: The
expansion rate of the universe is determined by the {\em total energy
density} in the universe by the relation given in the Friedman equation.
As can be seen from our dispersion function which approaches conventional
cosmology in the subplanckian regime ($k \le M_{pl}$), most of the other
contributions to the energy density are not frozen modes. Therefore the
Hubble rate $H^2$ is not always proportional to the dark energy of the
frozen modes due to the contributions in $H^2$ from other forms of energy
densitites.$H^2$ is dominated by frozen modes (and thus proportional
to the dark energy $\rho_{DE}$) only at some late times $t \ge t_E$ when
all other energy contributions $\rho_{other}$ have diluted enough below
$\rho_{DE}$ due to their redshift.

The quantitative effort we have made in this work suggests
that an interpretation of the dark energy
in terms of string theory is more convincing than either
a simple cosmological constant or the use of a slowly-
varying scalar field with fine tuned parameters.

\section*{Acknowledgements}

PHF acknowledges 
the support of the Office of High Energy, US Department
of Energy under Grant No. DE-FG02-97ER41036.

\end{document}